A GROSS-PITAEVSKII TREATMENT FOR SUPERSOLID HE

Philip W Anderson, Princeton University

ABSTRACT

**The observations of non-linear rotational susceptibility ("NCRI") in samples of solid He below 1-200 mK temperatures are conjectured to be describable in terms of a rarified Gross-Pitaevskii superfluid of vacancies (or, more generally, incommensuracies) with a transition temperature of about 50 mK, whose density is locally enhanced by crystal imperfections. We argue that the observations can be much affected by this density enhancement. We argue that it is likely that every pure Bose solid's ground state is supersolid.**

The Gross-Pitaevskii equation for the order parameter[1], and the corresponding free energy, has become standard for treating Bose condensation in cold atomic gases. The assumption on which this theory was based, essentially that the bosons be dilute relative to the range of their interaction, are well satisfied in these systems, which they are not for liquid HeII (and correspondingly there are complications in HeII, specifically the roton spectrum and the large depletion of the condensate.) One might think that the situation would be even worse in solid He, but in fact the opposite may be the case.

The solid is of course dense, but the experiments indicating supersolidity [2] show that the amount of matter that actually flows is a few percent of the atoms in reasonably well-frozen samples, and in good crystals much less than that. The earliest theoretical paper on supersolidity[3] proposed that it might consist of bose condensation of a small density of vacancies in the solid substrate, and the wave function which I have proposed as a heuristic description of the phenomena[4] can be interpreted in those terms. Reference [4] makes it clear that vacancy flow and substrate atom flow are essentially equivalent, so as a shorthand we will hereafter speak of vacancies and of a quantum (boson) field representing them. But I should say clearly that "vacancies" as used henceforth are a shorthand for the fluctuations in particle number which make it possible to define a phase of the particle field, which in the pure crystal can be equally matched between particles and

holes and are caused by overlap of local boson wave functions. (see Appendix)

Vacancies certainly interact repulsively as hard-core bosons, since no two can be on the same site; there may also be a long-range interaction due to elasticity which is thought to be attractive but probably doesn't play any dynamical role, that is simply changes the chemical potential. The vacancy density is, especially in a good crystal, very low, experimentally of order 3x $(10)^{-4}$.

The Gross-Pitaevskii free energy is

$$E = |(\hbar \nabla \Psi)|^2 / 2m^* + V(r) | \Psi |^2 + \frac{1}{2} g | \Psi |^4$$

and the G - P equation is

$$\mu \Psi = -\hbar^2 \nabla^2 \Psi / 2m^* + V(r) \Psi + g | \Psi |^2 \Psi \qquad [1]$$

Here $\Psi$ is the "order parameter" (to be thought of as the mean of the vacancy boson field), m* an effective mass, g a coupling constant $= h^2 a / \pi m^*$ (a is the scattering length, presumed repulsive), and $\mu$ the chemical potential for vacancies. V(r) is, in the gas case, simply the trap potential, but in our case it is included to take into account that dislocations, surfaces, grain boundaries, and possibly He-3 impurities are all attractive sites for vacancies. All of these parameters and their values need further discussion, especially since our physical conclusions depend crucially on them. [1] is essentially just a coarse-graining of the mean-field Schrodinger equation. The time-dependent equation which controls collective modes and vortex behavior is obtained by replacing $\mu$ by $\dfrac{\hbar \partial}{i \partial t}$.

The chemical potential tells us how many vacancies there are. The strong arguments of Reatto and Chester[5] based on a Bijl-Jastrow wave function for the ground state, as well as recent estimates by Reatto[6] and Boronat[7], are in favor of there being, even in the pure crystal, vacancies at the $10^{-4}$ level. (See Appendix for further discussion of this issue.) Repeated efforts by Chan and coworkers to grow perfect, pure single crystals have always observed non-classical rotational inertia (NCRI) on this level,[8] and this level or higher has been repeatedly confirmed by others[9]. (Simulations [10] using path-integral Monte Carlo have been strongly claimed to "prove" the nonexistence of vacancies, but among other difficulties the equivalent temperature in these simulations is well above the relevant temperature at which condensation takes place. In any case, simulating $10^4$ atoms well enough to find a single defect is beyond the capabilities of the methods.) We take the background density $|\Psi|^2$ relative to the solid for pure He4 to be $2-3(10)^{-4}$, which fixes $\mu$ in terms of g.

m* we assume to be fairly light relative to a helium atom. An estimate which is often quoted is $1/3 m_{He}$ and we will use that (another computer estimate is even smaller.[11]) This effective mass is such that the uncertainty energy necessary to localize it on a single site is of order 10°K. This is the same magnitude as estimates of the energy cost of a vacancy in ref 10, and suggests that those estimates may not have taken into account the kinetic energy which could be gained by delocalization. Regarding vacancies classically as strictly local configurations of the lattice is not reasonable.

m* and the density of the boson field allow an estimate of the superfluid transition temperature from the Bose-Einstein equation

$$kT_c = \frac{2\pi\hbar^2}{m}(\frac{N}{2.61V})^{2/3} \quad [2]$$

Putting in a typical solid density, the vacancy concentration of $2-3(10)^{-4}$, and a mass of $1/3 m_{He}$, one gets about 50-70 mdeg. This is very close to the transition temperature at which Kojima[12] and Chan et al[13] report thermal hysteresis in the NCRI. I have discussed elsewhere[14] why reversible NCRI appears so far above Tc. One expects true superflow to be observable only below this Tc, if at all.

The parameter g or, equivalently, the scattering length a, is not something one can estimate accurately. I would like to explore here the consequences of assuming that it is reasonably small. Perhaps one can justify this, again, from the fact that a light mass implies a somewhat extended lattice distortion. One may define a correlation length

$$\xi = 1/\sqrt{8\pi n_0 a} \text{ where } n_0 = |\Psi^2| \quad [3]$$

which is the exponential decay length of a small perturbation in the vacancy field, according to [1]. Given that $n_0=3 \times 10^{-4}$, even if a is a whole lattice constant ξ is 10 lattice constants or 3nm, and it would be reasonable for ξ to be an order of magnitude larger. This is still not quite the scale which Chan has shown[15] is showing up in the variation of NCRI with surface to volume ratio, but it's getting there.

What is of most interest is the effect of defects as attractive sites for vacancies. A dislocation core, for instance, is said to attract of the order of one vacancy per atomic length—calculated on the basis of localized, high-energy vacancies.[16] This amounts to a potential well in V which might be estimated as $V R^2 \approx$ 10-15°K, where R is its radius. Balancing this against the repulsive interaction $g\Psi^4$, it might be capable of attracting a cloud of $\propto 1/g$ vacancies with a radius of order ξ. Thus the effect of dislocations can be somewhat magnified. Correspondingly, one would expect there to be similar diffuse densities of delocalized vacancies around grain boundaries and near surfaces. I would consider this to be one of the few possibilities for explaining the degree to which crystal imperfection appears to enhance NCRI.

The question of the reason for the large effects of small concentrations of He-3 remains open.

In summary, it seems possible to provide an accounting of most of the puzzling properties of low-temperature solid He by describing it as a Gross-Pitaevskii fluid of delocalized quantum vacancies. Most successfully, the idea that the superfluid is an intrinsic property of the pure crystal which is locally enhanced by imperfections seems to account for the low and reasonably invariant genuine superfluid transition and the large variations in the quantity of superflow, which seem otherwise irreconcilable.

ACKNOWLEDGEMENTS

I would like to acknowledge the hospitality of the Fondation des Treilles for the workshop where this work was conceived, and the lectures and valuable discussions of M H W Chan, John Beamish, Luciano Reatto, Jean Dalibard, Jordi Boronat, and John Reppy at that workshop.

APPENDIX: PHYSICAL REASONS FOR THE SUPERSOLIDITY OF THE BOSE SOLID

As remarked in the paper, the argument of Chester-Reatto in ref [5], while not absolutely rigorous, seems adequate justification for assuming that there is a finite, if small, defect concentration in any Bose solid due to quantum fluctuations. Some additional considerations may throw some light on this question.

As described in a paper of Kohn[17], a very fundamental definition of a true insulator is that the energy is independent of the gauging of the particle field from site to site. For Fermions this is manifestly true of a band insulator since by Wannier's theorem a filled band is equivalent to occupying a local, orthonormal wave function at each site, and these can be gauged at will since the local fermions anticommute. (The Mott insulator is a more complicated case and was the point of Kohn's exercise.)

In the Bose case the Hartree-Fock approximation wave function for the ground state is [18]

$$\Psi = \prod_{sites\ i} b_i^* |vac\rangle; \quad b_i^* = \int d^3 r \varphi(r - r_i) \psi^*(r) \quad [A1]$$

here $\psi^*(r)$ creates a particle at point r in space, and the functions $\varphi$ are a set of functions exponentially localized to the neighborhood of site i, but everywhere positive,. They are positive everywhere because each is the ground state of a *different* local mean field equation and ground states have no nodes. Therefore, they are also necessarily not orthogonal. (The standard treatments[19] of the boson Hubbard model take the local functions as orthonormal, but they cannot be, even in an externally imposed lattice.)

Reference 17, and most further discussions, assume that the function (A1) is indeed a true "Mott" solid, in that apparently the sites can be gauged independently without doing anything but multiplying the function by the product of all the phase factors. But this is not correct; the operators $b_i^*$, $b_j$ do not commute. One might imagine that they could be orthogonalized, but this cannot be done without introducing nodes and thereby increasing the kinetic energy. Using the Wannier orthogonalization procedure is equivalent to uniformly filling a whole band, which clearly raises the energy. If they do not commute changing the relative phase of two of the b's causes phase-dependent changes in the density and therefore the energy. That is,

$$\text{If} \quad [b_i^*, b_j] = S_{ij} \text{ and } \Psi = \prod_i b_i^* |vac\rangle,$$

$$\langle n_i \rangle = 1 + \sum_j S_{ij} \qquad [A1]$$

Here $S_{ij}$ is dependent on the relative phases of the two bosons. Therefore the energy must also depend on relative phases, being lowest when the bosons are all in phase. This means that there are effectively both vacancies and interstitials in the simple Hartree wave function [A1].

We may schematize the phase dependent energy by an effective Hamiltonian,

$$H_{phase} = \sum_{ij} t_{ij} \cos(\phi_i - \phi_j), \text{ which implies that}$$

there is a supercurrent $J = t_{ij} \sin(\phi_i - \phi_j)$

between each pair of sites. These supercurrents are divergenceless; they represent an entirely separate dynamical system from motion of the lattice sites. It must be characterized by a system of vortices: a vortex fluid. Incidentally, it does not have off-diagonal long-range order in the usual sense, since number fluctuations are suppressed; but this is true of the superconductor as well.

It may not have escaped the reader that the above arguments apply to any Bose solid; but the amplitude of atom exchange is likely to be unmeasurably small in all cases but He (and possibly pure solid parahydrogen).

It remains to justify the use of a Hartree-Fock wave function even though the atoms are very strongly interacting. Every group who has tried to conceive of a trial function to describe the quantum solid has used the periodic product wave function and corrected for strong correlations with a Bijl-Jastrow or generalized Bijl-Jastrow product function, and I think of my procedure as referring to the underlying periodic solid wave function, corrected by some such procedure. Everything we have said applies to such a periodic wave function.